\documentclass{article}
\usepackage{ijcai21}

\usepackage{times}
\usepackage{soul}
\usepackage{url}
\usepackage[hidelinks]{hyperref}
\usepackage[utf8]{inputenc}
\usepackage[small]{caption}
\usepackage{graphicx}
\usepackage{amsmath}
\usepackage{multirow}
\usepackage{amsthm}
\usepackage{amsfonts}
\usepackage{booktabs}
\usepackage{algorithm}
\usepackage{color}
\usepackage{algorithmic}
\title{Electrocardio Panorama: Synthesizing New ECG Views with Self-supervision}

\author{
Jintai Chen$^1$\footnote{Co-first authors.} \and
Xiangshang Zheng$^{1*}$\and
Hongyun Yu$^{1*}$\and
Danny Z. Chen$^2$ \and
Jian Wu$^3$\footnote{The corresponding author.}
\affiliations
{\small
$^1$College of Computer Science and Technology, Zhejiang University, Hangzhou, China\\
$^2$ Department of Computer Science and Engineering, University of Notre Dame, Notre Dame, IN 46556, USA\\
$^3$ The First Affiliated Hospital, and Department of Public Health, Zhejiang University School of Medicine, Hangzhou, China}
\emails
{\small
jtchen721@gmail.com,
\{xszheng,yuhongyun777,wujian2000\}@zju.edu.cn,
dchen@nd.edu}
}

\begin{document}

\maketitle

\begin{abstract}
Multi-lead electrocardiogram (ECG) provides clinical information of heartbeats from several fixed viewpoints determined by the lead positioning. However, it is often not satisfactory to visualize ECG signals in these fixed and limited views, as some clinically useful information is represented only from a few specific ECG viewpoints. For the first time, we propose a new concept, Electrocardio Panorama, which allows visualizing ECG signals from any queried viewpoints. To build Electrocardio Panorama, we assume that an underlying electrocardio field exists, representing locations, magnitudes, and directions of ECG signals. We present a \textbf{N}eural \textbf{e}lectrocardio \textbf{f}ield \textbf{Net}work (Nef-Net), which first predicts the electrocardio field representation by using a sparse set of one or few input ECG views and then synthesizes Electrocardio Panorama based on the predicted representations. Specially, to better disentangle electrocardio field information from viewpoint biases, a new \textit{Angular Encoding} is proposed to process viewpoint angles. Also, we propose a self-supervised learning approach called \textit{Standin Learning}, which helps model the electrocardio field without direct supervision. Further, with very few modifications, Nef-Net can also synthesize ECG signals from scratch. Experiments verify that our Nef-Net performs well on Electrocardio Panorama synthesis, and outperforms the previous work on the auxiliary tasks (ECG view transformation and ECG synthesis from scratch). The codes and the division labels of cardiac cycles and ECG deflections on Tianchi ECG and PTB datasets are available at \url{https://github.com/WhatAShot/Electrocardio-Panorama}.
\end{abstract}
\section{Introduction}
\begin{figure}[hbt!]
    \centering
    \includegraphics[width=0.48\textwidth]{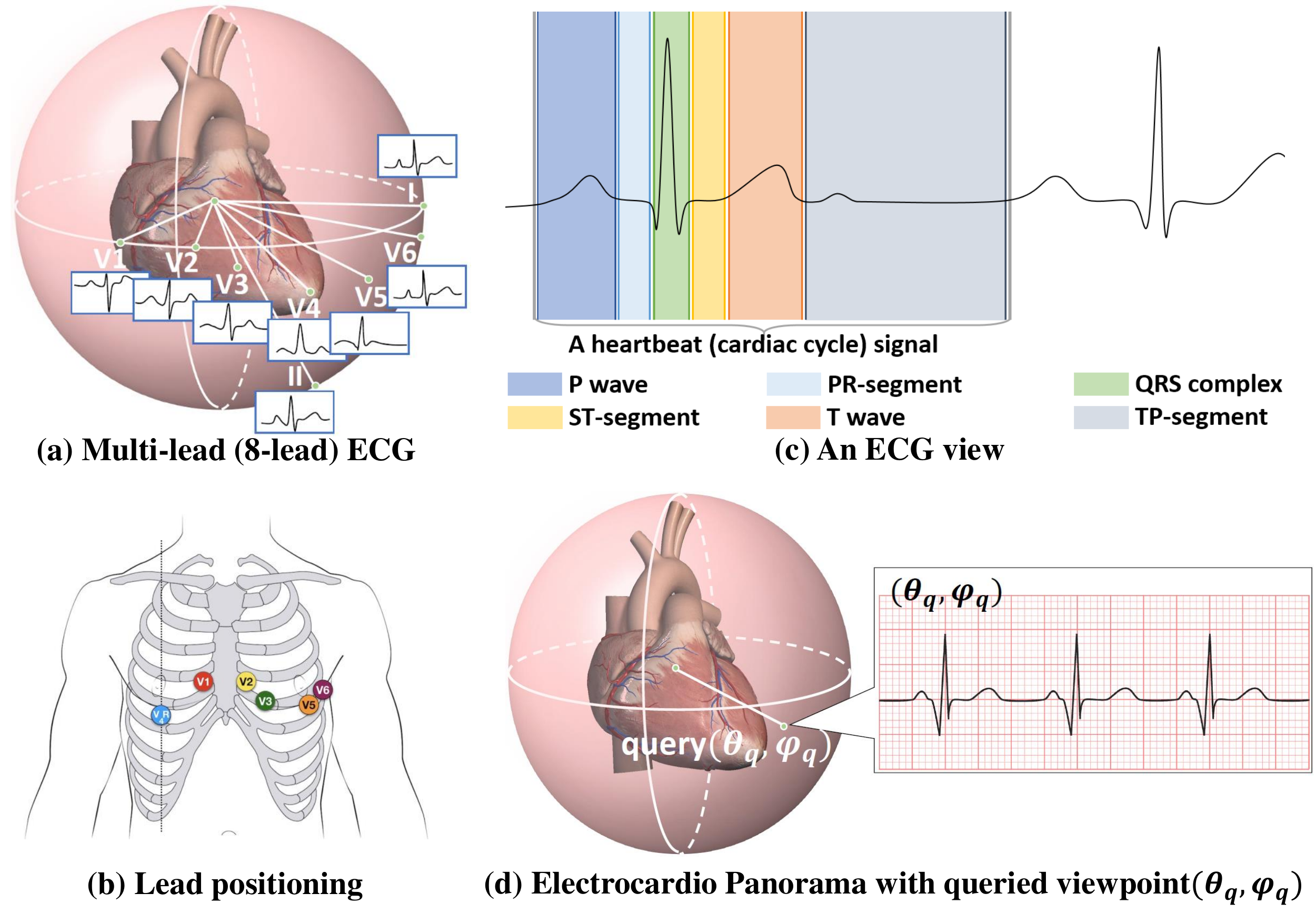}
    \caption{\textbf{(a)} A conventional multi-lead (e.g., 8-lead) ECG represents multi-view ECG signals from viewpoints determined by the lead positioning. \textbf{(b)} Illustrating the conventional lead positioning. \textbf{(c)} A continuous ECG signal can be divided into several cardiac cycles, and each cardiac cycle has 6 types of non-overlapping deflections, which are colored for better viewing. \textbf{(d)} Electrocardio Panorama can provide ECG signals from any queried viewpoints.}
    \label{fig:background}
\end{figure}
Electrocardiogram (ECG) has been instrumental in saving millions of lives since it emerged. Multi-lead ECG visualizes heartbeat signals in several views (see Fig.~\ref{fig:background}(a)), and the corresponding viewpoints are physically determined by the lead positioning (Fig.~\ref{fig:background}(b)\footnote{From \url{https://litfl.com/right-ventricular-infarction-ecg-library/} .}). In the past years, ECG visualization approaches have continued to evolve quickly. On one hand, more clinical information was recorded as the number of views increased from 3 to 8, 12, and even 18, with each view recording the signals (see Fig.~\ref{fig:background}(c)) from one specific viewpoint. For example, the increase of viewpoints helps determine the exact origin of the arrhythmia locating accuracy for the radiofrequency catheter ablation (RFCA) surgery. On the other hand, the visualization form has become increasingly more read-friendly~\cite{graybiel1946electrocardiography,case1979sequential}. Still, current ECG visualization approaches do not meet all the needs, as some clinical information is only intuitively represented in certain specific views (possibly outside of the conventional views).\\
\indent In this paper, we propose a novel concept, Electrocardio Panorama, which allows real-time querying of any ECG views, as illustrated in Fig.~\ref{fig:background}(d). Conventional ECG viewpoints are physically determined by the leads, while Electrocardio Panorama seeks to decouple the ECG views from the leads. Such panoramic representation increases the value of ECG monitoring, and thus is potentially life-saving.\\
\indent Some literature assumed an ECG view as the recording of an underlying electrocardio field from the corresponding viewpoint~\cite{grant1950spatial}, as illustrated in Fig.~\ref{fig:background}(a). Following this assumption, it is possible to synthesize Electrocardio Panorama based on the underlying electrocardio field. Motivated by NeRF~\cite{nerf}, we present a new neural network, called \textbf{N}eural \textbf{e}lectrocardio \textbf{f}ield \textbf{Net}work (Nef-Net), which encodes a sparse set of one or few input ECG views to predict the electrocardio field representation, and a reverse model (decoder) utilizes this representation to synthesize new ECG views. Specifically, we model the electrocardio field by using a basic representation and a set of deflection representations for separately modeling 6 ECG signal deflections (such 6 deflections are shown in Fig.~\ref{fig:background}(c)), as ECG deflections contain comprehensive personalized or morbid information. Since there is no direct supervision for electrocardio field modeling, we propose a novel \textit{Angular Encoding} approach to process viewpoint information (represented by angles), and a novel self-supervised learning approach called \textit{Standin Learning} to help disentangle electrocardio field information from viewpoint biases.\\
\indent Motivated by representation interpolations~\cite{stylegan}, our Nef-Net can also be used to synthesize ECG signals from scratch by mixing the extracted electrocardio field representations without explicitly modeling the latent distributions.  This work has five major \textbf{contributions}:\\
\indent \textbf{(A)} To our best knowledge, we are the first to propose the concept of Electrocardio Panorama, which decouples ECG signal recordings (views) from physical lead positioning and can provide ECG signals from any desired viewpoints.\\
\indent \textbf{(B)} We present a new \textit{Angular Encoding} to construct a high-dimensional angle space, which is helpful to disentangling electrocardio field information from viewpoint biases.\\
\indent \textbf{(C)} We propose a novel self-supervised learning approach, \textit{Standin Learning}, for electrocardio field information purification, which we verify to be effective.\\
\indent \textbf{(D)} Our proposed Nef-Net can synthesize the Electrocardio Panorama using sparse input ECG views, and can from scratch synthesize multi-lead ECG signals.\\
\indent \textbf{(E)} We publish cardiac cycle division and deflection division labels (e.g., see Fig.~\ref{fig:background}(c)) for two large open ECG datasets, contributing to the ECG research and community.
\section{Background and related work}
\paragraph{Electrocardiogram (ECG).} ECG signals record the heartbeats in multiple views, which are determined by the physical lead positioning~\cite{anderson1994panoramic} (see Fig.~\ref{fig:background}(a)-(b)). Currently, the most widely used ECG recording method is the standard 12-lead system, utilizing 12 ECG views (see Fig.~\ref{fig:background}(a)). A signal in a ECG view can be divided into several cardiac cycles (each represents a heartbeat), and each cardiac cycle signal can be divided into 6 non-overlapping deflections (containing rich personalized and morbid information): P wave, PR-segment, QRS complex, ST-segment, T wave, and TP-segment (see Fig.~\ref{fig:background}(c)). Besides, there is a U wave in the TP-segment, but it is often ignored in practice due to its minimal deflection.
\paragraph{ECG visualization.} Grant~\shortcite{grant1950spatial} indicated that the P wave and QRS complex were clinically useful and suggested to interpret ECG signals based on them. Interestingly, doctors summarized the correlations of various ECG views and proposed many visualization methods for multi-lead ECG in~\cite{graybiel1946electrocardiography,case1979sequential,anderson1994panoramic} based on their ECG reading preferences. These studies~\cite{case1979sequential,anderson1994panoramic} suggested that the limited and fixed ECG views were far from satisfactory, as each doctor has personal habit in ECG reading. Also, Anderson et al.~\shortcite{anderson1994panoramic} pointed out that some clinical information was more evident in some of the views. Hence, we propose a new concept, Electrocardio Panorama, which allows to provide ECG signals in any views.
\paragraph{ECG synthesis and view transformation.}\  Some prior work~\cite{golany2020improving,simgans} employed conditional generative adversarial networks (conditional GANs)~\cite{cgans} to synthesize personalized or morbid ECG signals. But, these methods could only synthesize signals of one view and did not consider the correlations among various views, thus giving only limited assistance in clinical practice. In~\cite{lee2019synthesis}, V-lead ECG signals (``V'' is a kind of ECG lead) were synthesized using ECG signals recorded by other leads. Earlier work~\cite{edenbrandt1988vectorcardiogram,kors1990reconstruction} explored ECG view transformation between the Frank lead~\cite{franklead} and standard lead system\footnote{The Frank is another ECG systems.}, and was meticulously compared in~\cite{comparison}. Recently, some neural network based regression models~\cite{lstm,vaecnn} were proposed for ECG view transformation. However, these methods were confined to the known ECG view transformation and could not synthesize new ECG views.
\section{Neural electrocardio field network (Nef-Net)}
\begin{figure*}[hbt]
    \centering
    \includegraphics[width=0.82\textwidth]{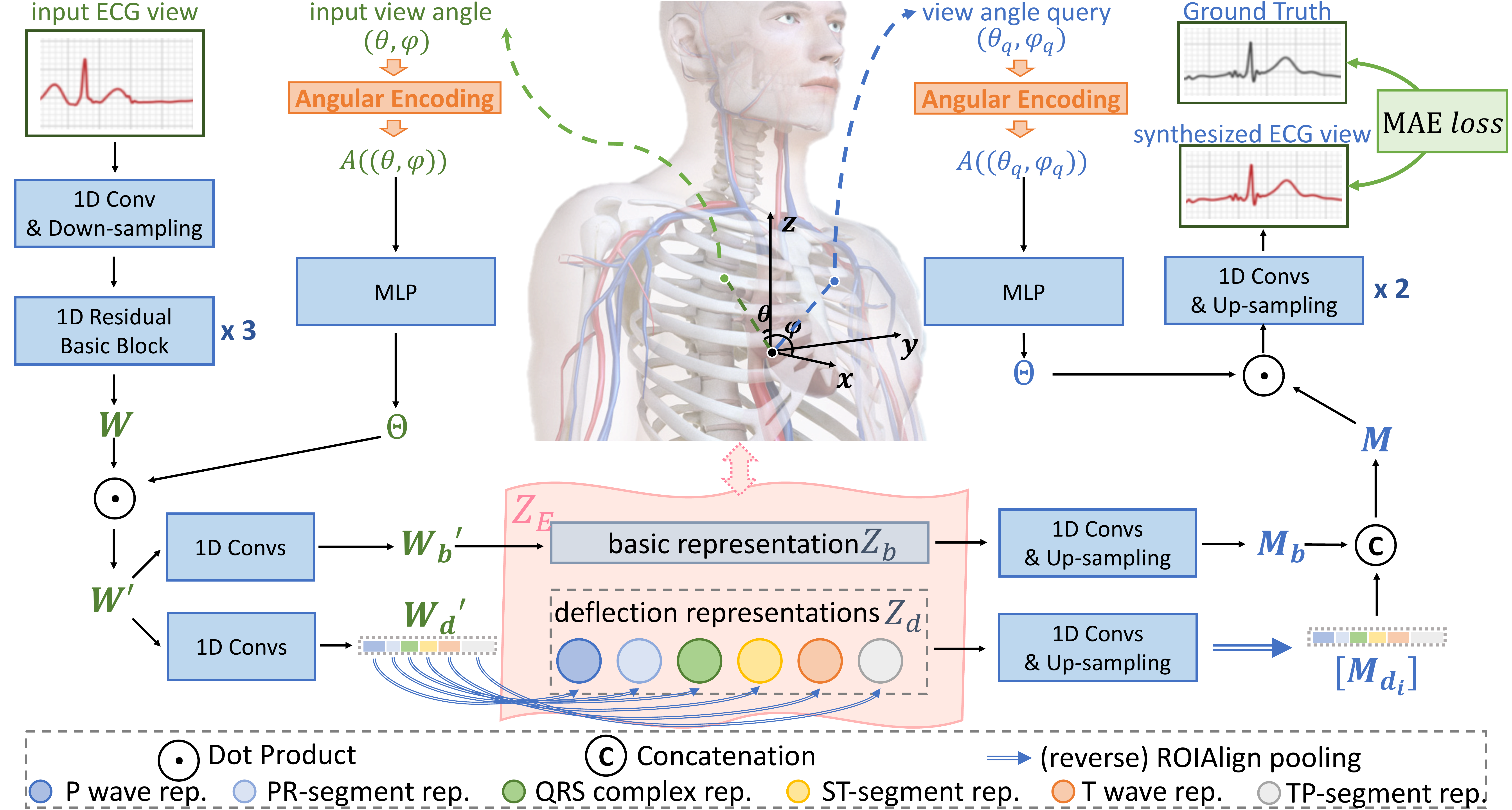}
    \caption{\textbf{Our proposed Nef-Net architecture for Electrocardio Panorama synthesis (illustrated using one input ECG view).} Nef-Net extracts the electrocardio field representation $Z_E$ from the input ECG views, and synthesizes ECG signals of a new view based on the learned representation $Z_E$ and the queried viewpoint $(\theta_q, \varphi_q)$. In the legend, ``rep.'' means ``representation''.}
    \label{fig:framework}
\end{figure*}
\subsection{Architecture}\label{seg:arc}
Nef-Net is based on an encoder-decoder architecture, as shown in Fig.~\ref{fig:framework}. In \textbf{encoding}, Nef-Net uses ECG signals in one or few views with the corresponding viewpoints to predict the electrocardio field representation; in \textbf{synthesizing}, a reverse model (decoder) uses this representation to predict new ECG views conditioned on the queried viewpoints. To explore viewpoint information, a new \textbf{\textit{Angular Encoding}} is used in both the encoding and synthesizing processes.
\paragraph{Input structure.} In our paper, a tuple containing one cardiac cycle view and its viewpoint is treated as one sample. We build a spherical coordinate system, with its origin point at the central electric terminal, the anatomical sagittal axis as the \textit{x-axis}, the inverse frontal axis as the \textit{y-axis}, and the vertical axis as the \textit{z-axis} (see Fig.~\ref{fig:framework}). Note that the radius is viewed as independent of ECG signals. Thus, we define a viewpoint by a polar angle $\theta$ and an azimuthal angle $\varphi$. In the following, we first depict the processing of a single input ECG view and then multiple input views. A single-view signal is denoted by $x$ of length $t$ (in time dimension) and the viewpoint is denoted by $(\theta, \varphi)$. A cardiac cycle signal with $L$ views is denoted by $\{x^{(l)}\}$ ($l \in \{1,2, \ldots, L\}$). In the decoder, a new view is synthesized conditioned on a queried viewpoint $(\theta^q, \varphi^q)$.
\paragraph{Encoding.} Given a cardiac cycle signal $x$ and the corresponding viewpoint $(\theta, \varphi)$, we first extract a time-aligned feature volume $W=f(x)$ by a 1D convolution module $f$, and extract a high dimensional angle feature $\Theta = g(A(\theta, \varphi))$ by a multi-layer perceptron (MLP) $g$ and a new \textit{Angular Encoding} process $A$, where $W \in \mathbb{R}^{c \times t'}, \Theta \in \mathbb{R}^{c}$, $c$ is the channel size, and $t'$ is the time dimension size. Then, we remove the viewpoint information from the feature volume $W$ by $W^\prime_{(c \times t')} = W_{(c \times t')} \odot \Theta^\prime_{(c \times t')}$, where $\Theta^\prime_{(c \times t')}$ is the stacking of $\Theta$ by $t'$ times and $\odot$ is the dot product.\\
\indent Since the deflections contain rich diagnostic features, we learn the electrocardio field representation, $Z_E$, using a basic representation $Z_b$ and a set of deflection representations $Z_{d_i}$ ($i\in \{1,2,3,4,5,6\}$ indexing the deflection types). The basic representation $Z_b$ learns the global information of the underlying electrocardio field, and each deflection representation $Z_d$ separately models each type of deflection. 1D convolution modules are then used to project $W^\prime$ to predict $W_b^\prime$ and $W_d^\prime$ for $Z_b$ and $Z_d$, respectively.  We project features of the $i$-th deflection $d_i$ on $W_d^\prime$ into the representation space $Z_{d_i}$ using ROIAlign pooling~\cite{he2017mask}, as the lengths of the same type of deflections vary among cardiac cycle signals but the size of $Z_{d_i}$ is fixed. The range (in $t^\prime$-dimension) of the deflection $d_i$'s features on $W_d^\prime$ is denoted by $\tau_{i} - \tau_{i-1}$, where $\tau_i= \frac{T_w}{T_x} \times D_i$, and $D_i$ is the demarcation point of the $(i-1)$-th and $i$-th deflections in the original signal $x$ ($D_i$ is given). $T_x$, $T_w$ are the lengths in time-dimension of $x$ and $W_d^\prime$, respectively. Specifically, $D_0=0$, $D_6 = T_x$, and $\tau_i$ is a value as in~\cite{fastrcnn} that varies among cardiac cycles.
\paragraph{Synthesizing.} New view synthesizing can be regarded as a reverse process of encoding, using $Z_E$ to predict ECG views. Each deflection representation $Z_{d_i}$ is first processed by a 1D convolution module (not shared among deflection representations) to obtain features $M^\prime_{d_i}$, and then is projected into features $M_{d_i}$ to restore the length proportions of the deflections using a reverse ROIAlign pooling (reROI):
\begin{equation}\label{reROI}
    M_{d}[:, \tau_{i-1}:\tau_i] = M_{d_i} = \text{reROI}(M^\prime_{d_i})
\end{equation}
where $\tau_i$ is computed as in encoding, and the exact grid value in $M_{d}$ is computed by linear interpolation like ROIAlign. We set $M_d$ and $M_b$ (from $Z_b$) to have an identical size. Then $M_b$ and $M_d$ are stacked along the channel dimension to form $M$. A queried viewpoint is processed by \textit{Augular Encoding} and an MLP, and is then multiplied by $M$ to add the queried viewpoint information. In this way, the features are processed to predict new views (as shown in Fig.~\ref{fig:framework}).
\paragraph{Angular Encoding.} As mentioned above, \textit{Angular Encoding} is used to handle the viewpoint angles in both encoding and decoding. Nef-Net learns to remove from or add viewpoint information to the electrocardio field (see Fig.~\ref{fig:framework}), but there is no direct supervision for the electrocardio field representation $Z_E$. Hence, ultimately, extracting viewpoint information is vital. It is intuitive to treat the viewpoint (angles) as a condition, and an ECG signal from a viewpoint ($\theta, \varphi$) is defined as $x = h(Z_E \ | \ (\theta, \varphi))$, where $h$ is a function (e.g., a decoder). Previous conditional GANs fed such conditions to a model directly. But, directly feeding raw angles may ignore the spatial relationships and thus result in poor performance. We propose a new \textit{Angular Encoding} to process viewpoint angles. We define a triplet as $\pi(\rho) = [\rho, \text{sin}(\rho), \text{cos}(\rho)]$ for an angle $\rho$. We define the \textit{Angular Encoding} $A$ by:
\begin{equation}
  A((\theta, \varphi))=A((\Tilde{\theta}, \Tilde{\varphi}))=[\pi(\Tilde{\theta}), \pi(\Tilde{\varphi}), \pi(\Tilde{\theta} + \Tilde{\varphi}), \pi(\Tilde{\theta} - \Tilde{\varphi})]
\end{equation}
where $\Tilde{\rho} = \rho + \epsilon$ and $\epsilon$ is a random variable sampled from a Gaussian distribution $\mathcal{N}(0, \frac{\pi}{50})$. \textit{Angular Encoding} transforms two angles into a 12-element tensor, which is then processed by an MLP. Since we treat the recorded ECG views as projections of the electrocardio field, some spatial projection related expressions (e.g., $\text{sin}(\theta)\text{cos}(\varphi)$) should be considered. However, it is hard to fit such expressions (especially multiplication) by using a finite MLP with raw angles, as an MLP can be regarded as an affine transformation if we ignore the activations. With $\text{sin}(\theta \pm \varphi)$ and $\text{cos}(\theta \pm \varphi)$, it is possible to represent the spatial projection related expressions via sum-to-product formulas. Thus, various relationships of angles (e.g., addition, cosine, multiplication) can be expressed with the \textit{Angular Encoding} and a finite MLP. We also add a perturbation $\epsilon$ to help avoid over-fitting the given angles, as Nef-Net is used to synthesize new views (of unknown patterns).
\subsection{Multi-view ECG signal processing}
Given a cardiac cycle signal $\{x^{(l)}\}$ ($l \in \{1,2, \ldots, L\}$) with $L$ views, we separately deal with these ECG views in parallel by group convolutions, instead of fusing them together as in the previous work~\cite{kachuee2018ecg,chen2020flow}. Formally, we predict $L$ basic representations and $L$ groups of deflection representations, and then compute the averaged basic representations $Z_b$ and averaged deflection representations $Z_d$ over the $L$ views as the input of the decoder.
\subsection{Learning}
To synthesize Electrocardio Panorama, we train Nef-Net under ECG view transformation supervision and use it for new view synthesis. A set of ECG data is partitioned for training and testing, and all kinds of views are divided into three groups, called the \textit{input group (IG)}, \textit{reconstruction group (RG)}, and \textit{synthesis group (SG)}. In training, Nef-Net takes the views in \textit{IG} of an ECG in the training set to reconstruct its views in \textit{RG}, under the guidance of the Mean Absolute Error (MAE). In testing, Nef-Net synthesizes views in \textit{SG} of an ECG in the testing set by using its views in \textit{IG}.
\paragraph{Standin Learning.} As mentioned above, Nef-Net is trained to synthesize new ECG views based on the electrocardio field. However, there is no direct supervision for the electrocardio field representation learning, and the input and output are related to the viewpoints. To deal with this, we perform self-supervision tricks~\cite{wk,fhz,MoCo} to eliminate the biases. To capture information from multiple views, a previous self-supervision work~\cite{tian2019contrastive} gathered view representations and implicitly built a scene embedding, by iteratively anchoring a view as the temporary target. But, this is ineffective as each view is somewhat biased and the intrinsic information is not directly extracted and fully used. We propose a new self-supervised learning approach called \textit{Standin Learning} for electrocardio field prediction, anchoring at the averaged representations as temporary targets to neutralize viewpoint biases. Given a cardiac cycle signal with $L$ views, $\{x^{(l)}\}$ ($l\in \{1,2, \ldots, L\} $), the corresponding basic representations and deflection representations are $\{Z^{(l)}_b\}$ and $\{Z^{(l)}_{d_i}\}$, and $Z_b=\sum^L_{l=1} Z^{(l)}_b / L$ and $Z_{d_i} = \sum^L_{l=1} Z^{(l)}_{d_i} / L$ are the averaged basic representation and averaged deflection representations, respectively. We define the contrastive loss for \textit{Standin Loss} $\mathcal{L}_Z=\frac{1}{2}(\mathcal{L}_{Z_b} + \mathcal{L}_{Z_d})$, where $\mathcal{L}_{Z_d}$ and $\mathcal{L}_{Z_d}$ are defined as:
\begin{equation}\label{eq:contrastive}
\left\{
\begin{aligned}
& \resizebox{0.9\hsize}{!}{$\mathcal{L}_{Z_b} = \sum\nolimits_{l=1}^L \mathcal{L}_{\text{MAE}}(h_{\text{sg}}(Z_b, Z_d|(\theta_q, \varphi_q)), h(Z^{(l)}_b, Z_d|(\theta_q, \varphi_q))) / L$}\\
& \resizebox{0.9\hsize}{!}{$\mathcal{L}_{Z_d} = \sum\nolimits_{l=1}^L \mathcal{L}_{\text{MAE}}(h_{\text{sg}}(Z_b, Z_d|(\theta_q, \varphi_q)), h(Z_b, Z^{(l)}_d|(\theta_q, \varphi_q))) / L$}
\end{aligned}
\right.
\end{equation}
where $h$ denotes the decoder and ``sg'' means ``stop gradient''. ``$\mathcal{L}_{\text{MAE}}(\text{target}, \text{prediction})$'' denotes the Mean Absolute Error. $\mathcal{L}_{Z}$ pushes the reconstruction based on the basic and deflection representations of one view close to that based on the averaged representations (fusing information of multiple views). We optimize the output rather than the representations to avoid changing the representation distribution and causing information loss. Hence, the electrocardio field representations are optimized iteratively: better single view representations improve the averaged representations, which further guide the single view representations to learn better.
\paragraph{Losses.} We supervise the Nef-Net training using the reconstruction loss MAE ($\mathcal{L}_{\text{MAE}}$) and contrastive loss $\mathcal{L}_Z$ as:
\begin{equation}
    \mathcal{L} = \mathcal{L}_{\text{MAE}}(x_q, \hat{x}_q) + \mathcal{L}_Z
\end{equation}
where $x_q$ and $\hat{x}_q = h(Z_b, Z_d|(\theta_q, \varphi_q))$ indicate the ground truth and predicted ECG signals of the queried view.
\section{Multi-lead ECG synthesis from scratch}
Our Nef-Net can also be utilized in ECG synthesis from scratch, without directly modeling the distributions as Generative Adversarial Nets (GAN)~\cite{gan} and Variational Auto-Encoder (VAE)~\cite{vae1,vae2}. Since ECG signal features are subtle, a simple representation distribution assumption (e.g., Gaussian distribution as in VAE) or a discriminator-guided distribution may not be suitable. Motivated by representation interpolations~\cite{stylegan} and Mixup~\cite{mixup,chen2020flow,chen2020transfer}, we synthesize electrocardio field representations of a new ECG by mixing the extracted electrocardio field representations of the same categories (e.g., diseases). As the range of each deflection type varies and the features of different types of deflections should not be mixed for new data synthesis, for this task we train a Nef-Net such that the electrocardio field representation is only predicted by the deflection representation learning branch (i.e., removing the branch for the basic representation). After training on a dataset, we introduce a memory bank \textbf{B} to store all the averaged deflection representations $Z_d$ for inference. Given the electrocardio field representations $Z^{(p)}_d, Z^{(q)}_d \in \textbf{B}$ of randomly selected cardiac cycles $p$ and $q$, the representation of a new cardiac cycle $n$ is:
\begin{equation}
    Z^{(n)}_d = a Z^{(p)}_d + (1-a) Z^{(q)}_d
\end{equation}
where $a \sim \text{Beta}(1.0, 1.0)$ as default. The new electrocardio field representation is then fed to the trained decoder $h_{\text{sg}}$ to synthesize new ECG data. As the deflection lengths $T_{d_i}$ of $p$ and $q$ may be different (used to compute $\tau_i$ for reverse ROIAlign pooling (Eq.~\ref{reROI})), we use the weighted average length for $n$, with $T^{(n)}_{d_i} = a \times T^{(p)}_{d_i} + (1-a) \times T^{(q)}_{d_i}$.
\section{Experiments}
\begin{figure}
    \centering
    \includegraphics[width=0.48\textwidth]{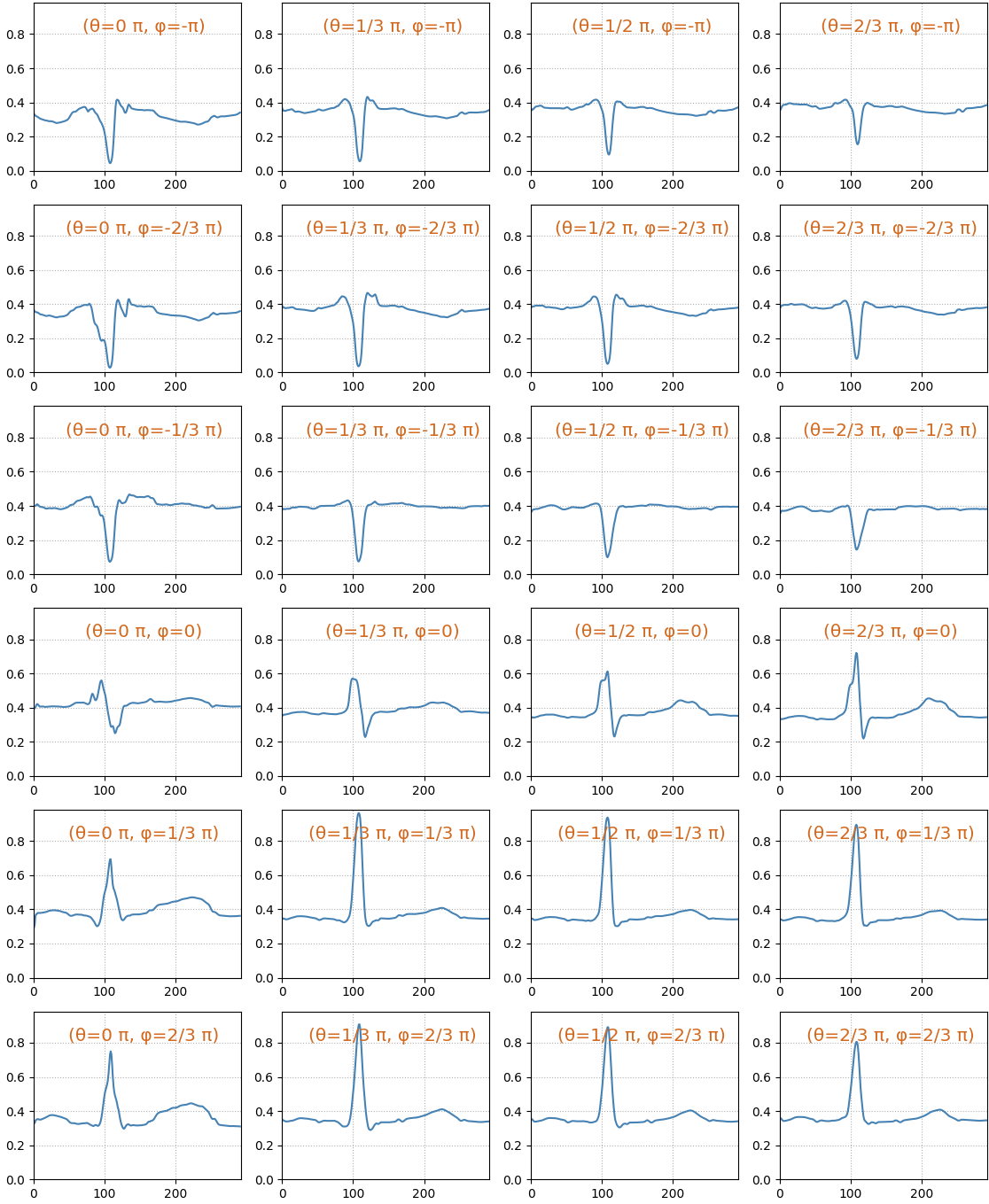}
    \caption{A case of Electrocardio Panorama.}
    \label{fig:vis_ep}
\end{figure}

\paragraph{Questions.} We conduct evaluations for five major questions in experiments. \textbf{(A)} Can Nef-Net synthesize Electrocardio Panorama, especially the ECG views that Nef-Net has not seen before? \textbf{(B)} As Nef-Net is trained with the view transformation supervision, can it outperform previous work in view transformation? \textbf{(C)} Are our proposed approaches (e.g., \textit{Angular Encoding}) helpful? \textbf{(D)} Is Nef-Net good at synthesizing ECG signals from scratch? \textbf{(E)} The model complexity of Nef-Net and its potential value in ECG monitoring?
\paragraph{Datasets.} We conduct experiments using the MIT-BIH dataset~\cite{PhysioNet}, PTB dataset~\cite{bousseljot1995nutzung}, and Tianchi ECG dataset\footnote{\url{https://tianchi.aliyun.com/competition/entrance/231754/information?lang=en-us}}. The MIT-BIH dataset contains 48 half-hour ECG signals recorded at a frequency of 360 Hertz, and each ECG signal is divided into several cardiac cycles (heartbeats). Each signal has 2 views, and the disease ground truth for each cardiac cycle is provided. The PTB dataset contains 549 12-lead ECG signals recorded at a frequency of 1,000 Hertz. The Tianchi dataset contains 31,779 12-lead ECG signals recorded at a frequency of 500 Hertz. Different from the MIT-BIH dataset, the PTB and Tianchi ECG datasets provide signals in 12 views. However, both these datasets do not provide the cardiac cycle division and the disease ground truth for cardiac cycles, and we annotate and publish the cardiac cycle division and deflection division ground truth for the PTB and Tianchi ECG datasets. For MIT-BIH, we divide the deflections using the package~\cite{makowski2016neurokit}. The PTB and Tianchi datasets are randomly partitioned into a training set and a test set with probabilities 0.8 and 0.2, respectively. The training-test partition for the MIT-BIH dataset follows~\cite{golany2020improving}, with 51,020 training samples and 49,711 test samples. 
We perform pre-processing, including de-noising (using the package~\cite{makowski2016neurokit}), interpolation to 500 Hertz, and linear scaling normalization to 0--1. To our best knowledge, no prior work is known on angle quantification for conventional ECG leads. Thus, we measure angles on 30 body CTs and compute the average angles for 12 leads (in lead-angle pair form), as: I-($\frac{\pi}{2}, \frac{\pi}{2}$), II-($\frac{5\pi}{6}, \frac{\pi}{2}$), III-($\frac{5\pi}{6}, -\frac{\pi}{2}$), aVR-($\frac{\pi}{3}, -\frac{\pi}{2}$), aVL-($\frac{\pi}{3}, \frac{\pi}{2}$), aVF-($\pi, \frac{\pi}{2}$), V1-($\frac{\pi}{2}, -\frac{\pi}{18}$), V2-($\frac{\pi}{2}, \frac{\pi}{18}$), V3-($\frac{19\pi}{36}, \frac{\pi}{12}$), V4-($\frac{11\pi}{20}, \frac{\pi}{6}$), V5-($\frac{8\pi}{15}, \frac{\pi}{3}$), V6-($\frac{8\pi}{15}, \frac{\pi}{2}$).
\paragraph{Experimental setups.} We use PyTorch 1.7.1 to implement Nef-Net. In training, the batch size is 32. Nef-Net is run 150 epochs in training. The learning rate is initialized to 0.1, and is reduced by $10 \times$ at the 50-th and 100-th epoch. We use SGD as the optimizer with momentum 0.9. We compute the structural similarity index measure (SSIM) and Peak Signal-to-Noise Ratio (PSNR) for evaluating the performances of new view synthesis and view transformation. Though SSIM is originally for 2D images, we regard ECG signal as a special case of images. We report the means and standard deviations over 3 runs with an RTX2080Ti GPU for all the experiments.
\paragraph{View transformation performances.} To examine the view transformation performances, we compare our Nef-Net with the known state-of-the-art models: conventional ECG lead transformation methods, including the Kors inverse matrix method (KIM)~\cite{kors1990reconstruction}, Kors quasi-orthogonal method (KQO)~\cite{kors1990reconstruction}, and Dower inverse matrix method~\cite{dower}; neural network based methods, including ensemble LSTM (E-LSTM)~\cite{lstm} and VAE-CNN~\cite{vaecnn}. The Dower, KQO, and KIM methods were originally for Frank vectorcardiogram reconstruction. We use them for view transformation with the procedure that we first transform the standard ECG signals to Frank vectorcardiogram, and then transform the vectorcardiogram back to the standard ECG signals. For fair comparison, the types of views used for input and the transformation (reconstruction) follow the previous work, and the numbers of views are orderly listed in the parentheses. As shown in Table~\ref{table:performance}, it is clear that our Nef-Net significantly outperforms the previous work on view transformation tasks, \textbf{by about 6--10 in PSNR and about 0.05--0.2 in SSIM}.
\begin{table}[t]
\centering
\caption{The performances (mean $\pm$ std) for view transformation tasks on the Tianchi and PTB datasets. The conventional methods (Dower, KQO, and KIM) did not show std. We report std only for PSNR, since the std of SSIM is too small. The numbers of views for input and transformation (reconstruction) are orderly listed in the parentheses, and the better performances are marked in \textbf{bold}.}\label{table:performance}
\resizebox{0.48\textwidth}{!}{
\begin{tabular}{lc|cc|cc}
\hline
\multicolumn{2}{c|}{\multirow{2}{*}{Methods}} & \multicolumn{2}{c|}{Tianchi}                 & \multicolumn{2}{c}{PTB}                    \\ \cline{3-6}
\multicolumn{2}{c|}{}                         & PSNR                 & SSIM                  & PSNR                 & SSIM                 \\ \hline
Dower                & (12, 12)              & 21.88                & 0.837                 & 19.99                & 0.844                \\
Nef-Net (ours)       & (12, 12)              & \multicolumn{1}{l}{\textbf{33.84}$\pm$0.30} & \multicolumn{1}{l|}{\textbf{0.973}} & \multicolumn{1}{l}{\textbf{30.74}$\pm$0.18} & \multicolumn{1}{l}{\textbf{0.983}} \\ \hline
KQO                  & (3, 12)               & 21.55                & 0.776                 & 19.86                & 0.810                \\
Nef-Net (ours)       & (3, 12)               & \multicolumn{1}{l}{\textbf{34.25}$\pm$0.31} & \multicolumn{1}{l|}{\textbf{0.972}} & \multicolumn{1}{l}{\textbf{31.53}$\pm$0.33} & \multicolumn{1}{l}{\textbf{0.975}} \\ \hline
KIM                  & (8, 12)               & 22.30                & 0.835                 & 21.91                & 0.862                \\
Nef-Net (ours)       & (8, 12)               & \multicolumn{1}{l}{\textbf{33.24}$\pm$0.29} & \multicolumn{1}{l|}{\textbf{0.973}} & \multicolumn{1}{l}{\textbf{30.46}$\pm$0.32} & \multicolumn{1}{l}{\textbf{0.982}} \\ \hline
VAE-CNN              & (1, 11)               & 26.40$\pm$0.21       & 0.918                 & 23.99$\pm$0.09       & 0.912                \\
Nef-Net (ours) & (1, 11) & \multicolumn{1}{l}{\textbf{32.94}$\pm$0.37} & \multicolumn{1}{l|}{\textbf{0.968}} & \multicolumn{1}{l}{\textbf{30.82}$\pm$0.07} & \multicolumn{1}{l}{\textbf{0.972}} \\ \hline
E-LSTM               & (3, 9)                & 22.88$\pm$0.33       & 0.830                 & 20.20$\pm$0.14       & 0.821                \\
Nef-Net (ours) & (3, 9)  & \multicolumn{1}{l}{\textbf{32.95}$\pm$0.22} & \multicolumn{1}{l|}{\textbf{0.971}} & \multicolumn{1}{l}{\textbf{30.48}$\pm$0.22} & \multicolumn{1}{l}{\textbf{0.971}} \\ \hline
\end{tabular}}
\end{table}
\begin{table}[t]
\centering
\caption{The performances for view synthesis on the Tianchi and PTB datasets. The numbers of views for input, transformation (reconstruction), and synthesis are orderly listed in parentheses. The part of views used to compute SSIM and PSNR is \underline{underlined}.}\label{table:synthesis}
\resizebox{0.48\textwidth}{!}{
\begin{tabular}{lc|cc|cc}
\hline
\multicolumn{2}{c|}{\multirow{2}{*}{Group}} & \multicolumn{2}{c|}{Tianchi} & \multicolumn{2}{c}{PTB} \\ \cline{3-6} 
\multicolumn{2}{c|}{}              & PSNR           & SSIM  & PSNR           & SSIM  \\ \hline
\multirow{2}{*}{(1) Nef-Net}           & (1, 9, \underline{2})   &29.60$\pm$0.35 & 0.949 & 28.95$\pm$0.65 & 0.951\\
                                   & (1, 9+\underline{2}, 0) & 31.74$\pm$0.21 & 0.953 & 29.77$\pm$0.06 & 0.952\\ \hline
\multirow{2}{*}{(2) Nef-Net} & (3, 7, \underline{2}) &29.19$\pm$0.19 & 0.943 & 29.27$\pm$0.26& 0.955\\
                                   & (3, 7+\underline{2}, 0) & 31.99$\pm$0.37 & 0.953 & 30.54$\pm$0.24 & 0.954 \\ \hline
\multirow{2}{*}{(3) Nef-Net} & (5, 5, \underline{2}) & 30.77$\pm$0.25 & 0.950 & 29.78$\pm$0.40 & 0.958\\
                                   & (5, 5+\underline{2}, 0) & 32.57$\pm$0.27 & 0.964 & 30.62$\pm$0.37 & 0.960 \\ \hline
\multirow{2}{*}{(4) Nef-Net}           & (1, 10, \underline{1}) & 29.97$\pm$0.36 & 0.938 & 29.31$\pm$0.05 & 0.954 \\
                                   & (1, 10+\underline{1}, 0) & 32.14$\pm$0.03 & 0.956 & 30.68$\pm$0.07 & 0.954\\ \hline
\multirow{2}{*}{(5) Nef-Net} & (3, 8, \underline{1}) & 30.51$\pm$0.45 & 0.946 & 29.43$\pm$0.25 & 0.951 \\
                                   & (3, 8+\underline{1}, 0) & 32.24$\pm$0.23 & 0.968 & 30.89$\pm$0.22 & 0.961 \\ \hline
\multirow{2}{*}{(6) Nef-Net} & (5, 6, \underline{1}) & 30.58$\pm$0.35 & 0.949 & 29.53$\pm$0.44 & 0.959\\
                                   & (5, 6+\underline{1}, 0) & 32.35$\pm$0.12 & 0.974 & 31.28$\pm$0.34 & 0.962\\ \hline
\end{tabular}}
\end{table}
\paragraph{New view synthesis performances.} Since we are the first to propose the Electrocardio Panorama, we compare the synthesized results with the transformed results of Nef-Net on the same types of views. As there are a large amount of combinations, we can only report some cases for references in Table~\ref{table:synthesis}. In Group (1) of Table~\ref{table:synthesis}, we train two Nef-Nets: one is trained with the 1 view input and with the 9 views as supervision, and synthesizes 2 views in testing; the other one is trained with the 1 input view, but with all the rest (11 views) as supervision, and predicts the corresponding 2 views (in test sets) in inference. The other groups are run following similar ways. We compare SSIM and PSNR on the 2 synthesized views of the test samples, and the key difference of these two models is whether the 2 types of views for comparison are used in training. We use only 1 or 2 views for synthesis, making the supervision similar. As shown in Table~\ref{table:synthesis}, the qualities of the synthesized views are comparable to the transformed views (in PSNR and SSIM). These results suggest that our synthesized views have high authenticity.\\
\indent Besides, comparing among the group of (1), (2), and (3), and among the group of (4), (5), and (6), one can see that with more input views, the synthesis performances often get better. Also, one may note that with one input view (groups (1) and (4)), the synthesis performances of Nef-Net are comparable to its transformation performances, and largely outperform the transformations of previous work (see Table~\ref{table:performance}). Thus, our Nef-Net is potentially useful in dynamic ECG applications (possibly with only one lead). \\
\indent To further verify the authenticity of the synthesized views, we also present one case of panoramic representations where the viewpoint angles $\theta$ and $\varphi$ are evenly distributed in the angle ranges with an interval $\frac{\pi}{3}$, as shown in Fig.~\ref{fig:vis_ep}. The synthesized ECG views are though to be highly trusty. 
\begin{table}[t]
\centering
\caption{Ablation study using the Tianchi ECG dataset.}\label{table:ab}
\resizebox{0.48\textwidth}{!}{
\begin{tabular}{l|ccccccc|cc|cc}
\hline
\multicolumn{1}{c|}{\multirow{2}{*}{}} &
  \multirow{2}{*}{$\epsilon$} & \multirow{2}{*}{\textit{AE}} & \multirow{2}{*}{$Z_d$} &  \multirow{2}{*}{$\mathcal{L}_{Z_b}$} & \multirow{2}{*}{$\mathcal{L}_{Z_d}$} &  \multirow{2}{*}{$\mathcal{L}^\prime_{Z_b}$} & \multirow{2}{*}{$\mathcal{L}^\prime_{Z_d}$} &  \multicolumn{2}{c|}{synthesis} & \multicolumn{2}{c}{transformation} \\ \cline{9-12} 
\multicolumn{1}{c|}{}             &   &   &   &   &   &  &  & PSNR & SSIM & PSNR & SSIM \\ \hline
(a)      & \checkmark & \checkmark & \checkmark & \checkmark & \checkmark &  &  & 29.34 & 0.957 & 32.32 & 0.964 \\ \hline
\multirow{2}{*}{(b)} &   & \checkmark & \checkmark & \checkmark & \checkmark &  &  &  29.18 &  0.945 & 31.90 &  0.955   \\
                                  & \checkmark &   & \checkmark & \checkmark & \checkmark &  &  &  28.44 &  0.941 &  32.35  &  0.964   \\ \hline
(c)          & \checkmark & \checkmark &   & \checkmark  &   &  &  &  28.64 & 0.946 &   31.77 & 0.959  \\ \hline
\multirow{3}{*}{(d)}             & \checkmark & \checkmark & \checkmark & \checkmark &   &   &  &  29.16 & 0.946 &  32.31 &  0.956 \\
                                  & \checkmark & \checkmark & \checkmark &   & \checkmark &  &  & 28.65 &  0.943 &  31.13 &   0.961  \\
                                  & \checkmark & \checkmark & \checkmark &   &   &  &  &  28.50 & 0.945 &  30.00 &  0.959 \\ \hline
\multirow{3}{*}{(e)}        & \checkmark & \checkmark & \checkmark &  & \checkmark & \checkmark &  & 27.02  & 0.936 & 27.20 &  0.936  \\
                                  & \checkmark & \checkmark & \checkmark & \checkmark &  &  & \checkmark &  28.91    &  0.934 &  32.07    &  0.948  \\
                                  & \checkmark & \checkmark & \checkmark & & & \checkmark  &  \checkmark  &  25.97    &   0.922 &  26.44   &  0.929   \\ \hline
\end{tabular}}
\end{table}
\begin{table}[t]
\centering
\caption{Classification performances (ROC-AUC) on the MIT-BIH dataset augmented by synthesized data. The best results are in \textbf{bold}.}\label{table:syn}
\resizebox{0.5\textwidth}{!}{
\begin{tabular}{cccccc}
\hline
Diseases & VGAN & DCGAN & SimVGAN & SimDCGAN & Nef-Net (ours) \\ \hline
SVEB & 0.768$\pm$0.06 & 0.685$\pm$0.04 & 0.765$\pm$0.07 & 0.724$\pm$0.05 & \textbf{0.782}$\pm$0.05 \\
VEB & 0.982$\pm$0.01 & 0.981$\pm$0.00 & 0.979$\pm$0.01 & 0.980$\pm$0.01 & \textbf{0.983}$\pm$0.00 \\
FUSION & 0.811$\pm$0.08 & 0.823$\pm$0.09 & 0.778$\pm$0.05 & 0.827$\pm$0.09 & \textbf{0.855}$\pm$0.07 \\ \hline
\end{tabular}}
\end{table}
\paragraph{Ablation study.} Based on the benchmark Nef-Net (4, 4, 4) (denoted by (a) in Table \ref{table:ab}), we evaluate the effects of our proposed approaches, including (b) \textit{Angular Encoding (AE)} and the perturbation $\epsilon$ used in \textit{AE}, (c) deflection modeling by independent representations $Z_d$, and (d) contrastive losses $\mathcal{L}_{Z_b}$ and $\mathcal{L}_{Z_d}$. Besides, to verify (e) the effectiveness of using averaged representations as the temporary targets for the contrastive loss, we use $\mathcal{L}^\prime_{Z_b}$ and $\mathcal{L}^\prime_{Z_b}$ to replace $\mathcal{L}_{Z_b}$ and $\mathcal{L}_{Z_d}$ in Eq.~(\ref{eq:contrastive}), respectively. For $\mathcal{L}^\prime_{Z_b}$, we replace the averaged basic representation $Z_b$ in $\mathcal{L}_{Z_b}$ with the basic representation of a random view (sampled once per iteration). For $\mathcal{L}^\prime_{Z_d}$, we replace the averaged deflection representations $Z_d$ in $\mathcal{L}_{Z_d}$ with the deflection representations of a random view. As shown in Table~\ref{table:ab}, all our proposed approaches are helpful in synthesis. One may notice that the models without \textit{Angular Encoding} or deflection representations can attain comparable performances for the view transformation tasks, but are not effective for new view synthesis. The results suggest that our \textit{Angular Encoding} and deflection modeling can help enhance generalization of models. As shown in (d) and (e) of Table~\ref{table:ab}, $\mathcal{L}_{Z}$ (especially $\mathcal{L}_{Z_b}$) and \textit{Standin Learning} help Nef-Net considerably on both the view transformation and view synthesis tasks.
\begin{figure}[t]
    \centering
    \includegraphics[width=0.48\textwidth]{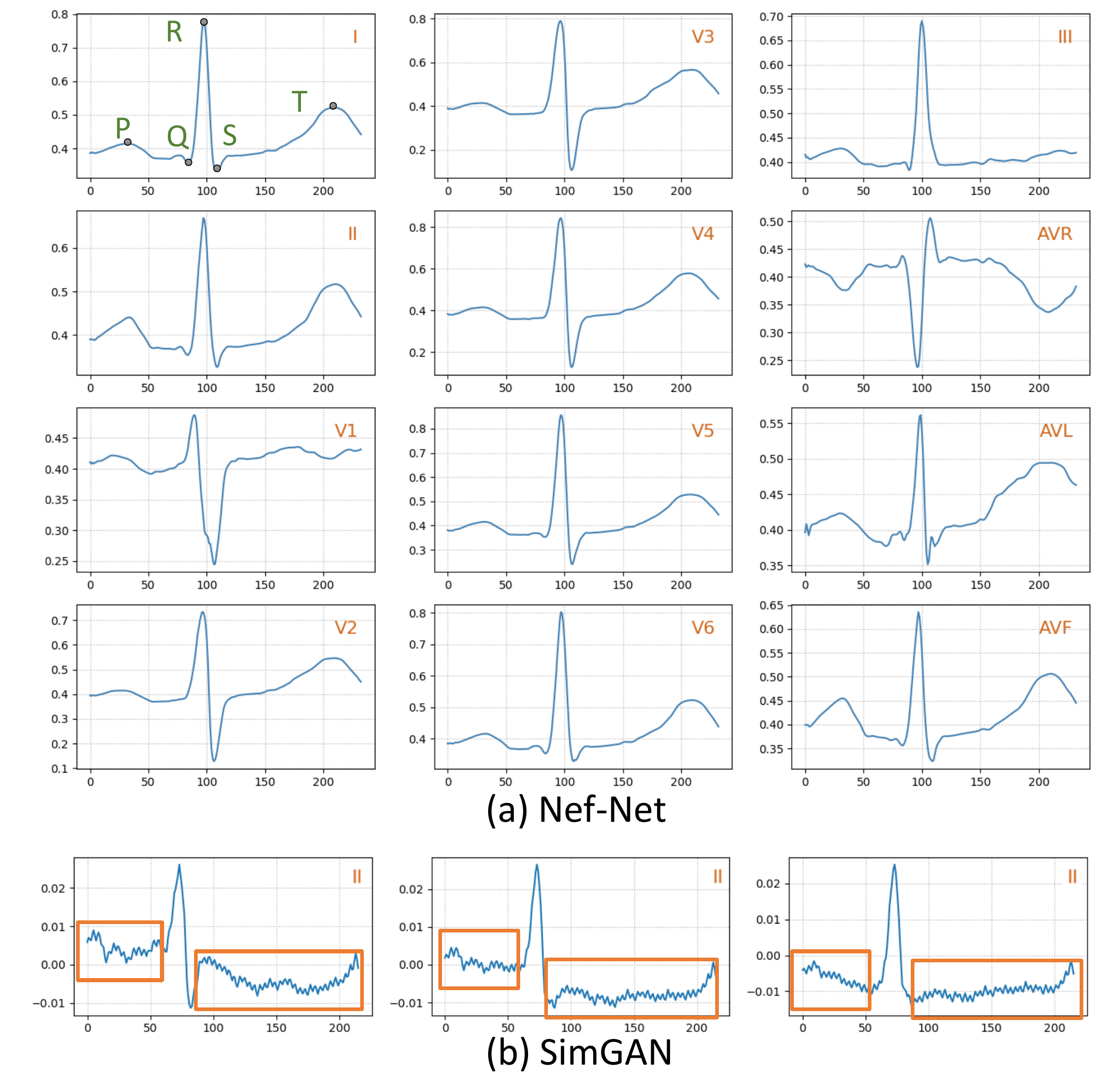}
    \caption{Examples of a multi-lead ECG synthesized from scratch by Nef-Net, and 3 single-lead ECG signals synthesized by SimGAN.}\label{fig:vis2}
\end{figure}
\paragraph{ECG synthesis from scratch.} To evaluate Nef-Net's capability to synthesize ECG signals from scratch, we compare its classification performances (with ROC-AUC) on morbid ECG signal synthesis with the state-of-the-art methods, simDCGAN and simVGAN~\cite{simgans}, DCGAN~\cite{dcgan}, and vanilla GAN~\cite{gan}, on the MIT-BIH dataset. The performances of these known methods are obtained by running the open source~\cite{simgans}, and the classifier for evaluating the synthesized samples is implemented as in~\cite{kachuee2018ecg}. All the other settings follow~\cite{simgans} for comparison. As shown in Table~\ref{table:syn}, our Nef-Net outperforms the previous work, our method outperforms the previous work by clear margins on those performance-unsaturated diseases: SVEB (1.4\%) and FUSION (2.8\%). For VEB, since the performances were close to 100\% in the previous work, there is not much room for improvement. Such performances reveal that our Nef-Net captures the key information of ECG.\\
\indent A visualization of multi-lead ECG data synthesis from scratch by Nef-Net is given in Fig.~\ref{fig:vis2}, which is considered as having a high degree of authenticity. As shown in Fig.~\ref{fig:vis2}, Nef-Net can synthesize multi-lead ECG signals from scratch, presenting very detailed waves and segments (e.g., the key waves are marked in the top-left sub-figure of Fig.~\ref{fig:vis2}(a)), while SimGAN can synthesize only single-lead ECG whose key waves are lost (e.g., P waves and T waves; see Fig.~\ref{fig:vis2}(b)). Also, the signals synthesized by SimGAN have considerable noise (e.g., as highlighted in the orange rectangles), while those synthesized by ours are nearly noise-free.
\paragraph{Model complexity.} In inference, the fps (frames per second) of our Nef-Net is 13.9, which can be used for real-time Electrocardio Panorama synthesis. The Nef-Net model size is 7.18 MB, and is feasible to be used on some mobile devices. Thus, Nef-Net is potentially helpful to medical doctors (e.g., for dynamic ECG (holter ECG) and home-ECG monitoring).
\section{Conclusions and Future Work}
In this paper, we proposed a new concept, Electrocardio Panorama, which decouples the ECG views from the physical lead positioning and can provide ECG signals of any queried views in real time. We presented a versatile neural network called Nef-Net, which can transform ECG views, synthesize Electrocardio Panorama, and synthesize ECG signals from scratch. Experiments verified that Nef-Net performs well on these tasks. To extract the electrocardio field representations without supervision, we proposed a new \textit{Angular Encoding} for angle processing and an effective self-supervised learning approach (\textit{Standin Learning}) to help extract and purify electrocardio field representations. Ablation study showed that our new approaches are helpful. Finally, the cardiac cycle and deflection division labels for the PTB and Tianchi ECG datasets are available for reference.\\
\indent There are some potential improvements for future research on Electrocardio Panorama. First, ECG signals are currently normalized before feeding into Nef-Net. However, normalization might neglect absolute amplitudes, which are important in practice. Second, an interesting issue is to consider ECG synthesis from scratch. Our Nef-Net can synthesize ECG data from scratch and significantly outperform the previous work. It is worth exploring to improve the GAN and VAE methods for ECG by incorporating Nef-Net. Third, although the datasets used in our study contain some ECG signals of patients with heart diseases, we did not explicitly utilize some specific operations for processing morbid ECG signals. Thus, it is important to investigate improving Electrocardio Panorama with respect to morbid ECG signals. Finally, developing a modern visualization system for Electrocardio Panorama is still an open research topic.
\section*{Acknowledgments}
This research was partially supported by the National Key Research and Development Program of China under grant No. 2019YFC0118802, the National Natural Science Foundation of China under grant No. 61672453, the Zhejiang University Education Foundation under grants No. K18-511120-004, and No. K17-511120-017, the Zhejiang public welfare technology research project under grant No. LGF20F020013, and the Key Laboratory of Medical Neurobiology of Zhejiang Province. D.Z. Chen’s research was supported in part by NSF Grant CCF-1617735.
\bibliographystyle{named}
\bibliography{myref}

\end{document}